\documentclass[prb,twocolumn]{revtex4}
\usepackage{bm}
\usepackage[dvips]{graphicx}

\newcommand*{\be}{\begin{equation}}
\newcommand*{\ee}{\end{equation}}
\newcommand*{\bea}{\begin{eqnarray}}
\newcommand*{\eea}{\end{eqnarray}}

\newcommand*{\sd}{^{\dagger}}

\begin{document}


\title{
Possible $Z_2$ phase and spin-charge separation\\
in electron-doped cuprate superconductors
}

\author{Tiago C. Ribeiro}
\author{Xiao-Gang Wen}
\affiliation{Department of Physics, Massachusetts Institute of Technology, Cambridge, Massachusetts 02139, USA}

\date{\today}

\begin{abstract}
The $SU(2)$ slave-boson mean-field theory for the $tt'J$ model is analyzed.
The role of next-nearest-neighbor hopping $t'$ on the phase diagram is
studied.  We find a pseudogap phase in hole-doped (HD) materials (where
$t'<0$). The pseudogap phase is a $U(1)$ spin liquid (the staggered-flux
phase) with a $U(1)$ gauge
interaction and no fractionalization.  This agrees with experiments on HD
samples. The same calculation also indicates that a positive $t'$ 
favors a $Z_2$ state with true spin-charge separation.  The $Z_{2}$ state that
exists when $t'\gtrsim 0.5J$ can be a candidate for the pseudogap phase of
electron-doped (ED) cuprates (if such a phase exists).  The experimental
situation in ED materials is also addressed.  
\end{abstract}


\maketitle

\section{\label{sec:intro}Introduction}

A complete understanding of the behavior displayed by high temperature
superconducting cuprates is still lacking.  In these materials, d-wave
superconducting (dSC) samples are obtained upon doping the parent compounds.
The latter are Heisenberg antiferromagnets. The presence of low energy
properties consistent with standard BCS theory in dSC samples is also
generally undisputed.  However, the phase-space region intervening between the
two aforementioned regimes hosts an unusual phenomenology -- and a quite
debated one as well.
In this paper, we will adopt the point of view that the unconventional 
behavior in underdoped cuprates is evidence of doped Mott insulator physics. 
To capture the important properties of these materials, which are then 
controlled by the large Mott gap and the two-dimensional spin 
interactions in the 
copper-oxide planes \cite{A8796}, we will use the slave-boson approach 
developed in Refs. \onlinecite{SLAVE_BOSON} and \onlinecite{WENLEENN}.

In underdoped electron-doped (ED) materials the antiferromagnetic (AF) 
phase is very robust. 
However that is not the case in their hole-doped (HD) counterparts.  
Instead, a pseudogap metallic normal
state appears in underdoped samples.  This is a paramagnetic state with a
$d_{x^2-y^2}$ gaped spectral function.  Reconciling the gap with the strong
local AF fluctuations is not trivial though. \cite{RW_SPIN}
One way to understand it is offered by the slave-boson approach to the $tJ$
model.  In particular, in Ref. \onlinecite{WENLEENN} a translation symmetric
pseudogap metallic state (the staggered-flux or s-flux state) was proposed.
It contains spinons and holons interacting via a $U(1)$ gauge field.  
This state was shown to bear enhanced AF fluctuations compatible with
experiments.\cite{RW_SPIN} The $U(1)$ gauge fluctuations play a key role in
the AF instability at very low doping as well as in the destruction of
quasiparticle peaks.\cite{RW_SPECTRAL,FT0103}  
However, as discussed in Ref. \onlinecite{RW_SPIN}, the
appearance of the spin-pseudogap below a certain energy and the existence of
well defined electronlike quasiparticles in the nodal direction above $T_c$
suggest that spin and charge recombine before the s-flux state reaches
the dSC state.

There is a competing scenario to describe this crossover from the s-flux to the
dSC phase which involves the emergence of an intervening state with a $Z_2$
gauge structure.\cite{SF0050} This state would result from the condensation of
spinon bilinears.\cite{RS9173,WENsrvb,WEN} The dSC state would then follow
from holon condensation.  $Z_2$ spin liquids are very interesting states
with true spin-charge separation.  They were predicted theoretically over
ten years ago.\cite{RS9173,WENsrvb} Yet they lack experimental
realization. The present work proposes to study the conditions favoring the
emergence of these spin liquid states as well as to discuss their relevance
for the cuprate superconductors.

Despite the effort to check the applicability of $Z_{2}$ spin liquids in the
context of high $T_{c}$ superconductivity, so far all the experimental tests
gave negative results.\cite{Z2CHECK_EX, Z2CHECK_TH} However, this does not
imply that $Z_2$ states do not exist in high $T_{c}$ samples. In this paper
we find that $Z_2$ states are indeed unlikely to exist in HD compounds. But a
$Z_2$-linear state (\textit{i.e.} with linear gapless spinon excitations) is
likely to appear as the pseudogap metallic state for ED materials.  
In superconducting (SC) samples without AF order spin liquid pseudogap 
signatures may emerge once SC coherence is
destroyed by thermal fluctuations or an external magnetic field.  
In the first case such a behavior is still to be reported.
Experimental evidence in the magnetic field driven normal state, however,
seems to be consistent with $Z_2$ pseudogap phenomenology.
\cite{KLEEFISCH_BISWAS_ALFF,HILL}

The $tJ$ model has a particle-hole symmetry.  In order to break it
we introduce the next-nearest-neighbor (NNN) hopping $t'$ term.
Taking both signs of $t'$ covers both ED ($t'>0$) and HD ($t'<0$)
cases.  In Sec. \ref{sec:su2_mf} we introduce the  $SU(2)$ MF theory for the
$tt'J$ model.  In Sec. \ref{sec:p_d} we present the resulting MF
phase diagrams.  For $t'<0$ our calculations are consistent with the previous
studies for $t'=0$ where only the s-flux state was obtained. However we find
that NNN hopping can stabilize a $Z_2$-linear state in the $tt'J$ model for
values of $t' \gtrsim 0.5J$.  These results show that unfrustrated hopping
favors fractionalized phases (Sec. \ref{sec:t'}).  In 
Sec. \ref{sec:application}
we establish the link between our results and some properties of ED cuprates.
We draw our main conclusions in Sec. \ref{sec:conclusion}.

\section{\label{sec:su2_mf} $SU(2)$ MF theory for the $tt'J$ model}

The $tt'J$ model is given by the Hamiltonian 
\begin{equation}
H_{tt'J} = H_{hf} + H_{hop} + H' 
\end{equation}
with 
\begin{eqnarray}
H_{hf} &=& \sum_{<ij>}J_{ij}(\vec{S_{i}}.\vec{S_{j}}-\frac{1}{4}),
\nonumber\\
H_{hop} &=& - \sum_{<ij>}t_{ij}\bm{P}(c_{i}\sd c_{j} + c_{j}\sd c_{i})\bm{P},
\nonumber\\
H' &=& \frac{1}{4} \sum_{<ij>}J_{ij} \left( 1-n_{i}n_{j} \right),
\end{eqnarray}
where sums are taken over pairs of sites $\big<ij\big>$. The exchange coupling
$J_{ij}$ is $J$ for $\big<ij\big>$ nearest-neighbor (NN) sites 
while the hopping parameter $t_{ij}$ equals $t$ for $\big<ij\big>$ NN sites,
$t'$ for $\big<ij\big>$ NNN sites. 
$\vec{S_{i}}$ is the spin on site $i$, $c_{i}\sd$ and $n_{i} = c_{i}\sd c_{i}$
are electron creation and occupation number operators and $\bm{P}$ projects
out doubly occupied sites.

To deal with the no double occupancy constraint the electron operator is
decoupled according to $c_{i\uparrow} = \frac{1}{\sqrt{2}} (
\psi_{i1}b_{i1}\sd+\psi_{i2}b_{i2}\sd )$ and $c_{i\downarrow} =
\frac{1}{\sqrt{2}} ( \psi_{i2}\sd b_{i1}\sd-\psi_{i1}\sd b_{i2}\sd )$ and the
spin operators are expressed as  $\vec{S_{i}} = \frac{1}{2} f_{i\alpha}\sd
\vec{\tau}_{\alpha \beta} f_{i\beta}$
where the $SU(2)$ doublets $\Psi_{i}\sd = [ \psi_{i1}\sd \psi_{i2}\sd ] = [
f_{i1}\sd f_{i2} ]$ and $h_{i}\sd = [ b_{i1}\sd b_{i2}\sd ]$ represent spinons
(fermionic particles carrying spin $\frac{1}{2}$) and holons (bosonic
particles carrying unit charge) respectively and $\bm{\vec{\tau}}$ are the
Pauli matrices.  These operators were introduced in Ref. \onlinecite{WENLEENN} to keep
the $SU(2)$ gauge structure of $H_{tt'J}$ even away from half-filling. 
Parton operators introduce unphysical degrees of freedom which are projected
out by implementing the constraint $\vec{Q}_{i} \equiv \Psi_{i}\sd \vec{\tau}
\Psi_{i} + h_{i}\sd \vec{\tau} h_{i} = 0$ (\textit{i.e.} physical states are
$SU(2)$ singlets on every site).  Introducing a Lagrange multiplier per
lattice site ($\vec{a}_{0i}$) the problem is reduced to one of lattice gauge
theory.\cite{WENLEENN}
Performing a Hartree-Fock type of decoupling of $H_{hf}$ and $H_{hop}$ 
(neglecting $H'$ for the time being)
and implementing the projection constraint at MF level only, the following MF
Hamiltonian is obtained:
\bea
H_{MF} &=&  \frac{3}{16} \sum_{<ij>} J_{ij} Tr\left(U_{ij} U_{ji} \right) 
+ \frac{1}{2} \sum_{<ij>} t_{ij} Tr\left(U_{ij} B_{ji} \right) \nonumber \\
&+& \sum_{i} \vec{a}_{0i} . \left( \Psi_{i}\sd \vec{\tau} \Psi_{i} + h_{i}\sd 
\vec{\tau} h_{i} \right) - \mu_{b} \sum_{i} h_{i}\sd h_{i} \nonumber \\ 
&-& \frac{1}{2} \sum_{<ij>} t_{ij} \left( h_{i}\sd U_{ij} h_{j} + h.c. \right) 
\nonumber \\
&-& \sum_{<ij>} \left[ \Psi_{i}\sd \left( \frac{3}{8} J_{ij} U_{ij} + 
\frac{1}{2} t_{ij} B_{ij} \right) \Psi_{j} + h.c. \right]
\label{eq:ham_MF}
\eea 
\noindent
where
$\mu_{b}$ is the chemical potential of holons.

Despite ignoring gauge fluctuations, we expect MF theory to account for 
some qualitatively right features.
Indeed, we have the following: 
\\ \noindent
\hspace*{1ex}\textit{i)} 
The energetics of the model is correctly captured at MF level.  This fact is
suggested by the $SU(2)$ slave-boson MF theory phase diagram for the $tJ$
model -- it includes the pseudogap regime (s-flux state)
that turns into the dSC state as holons become coherent, as well as the
strange-metal and Fermi-liquid regimes.\cite{WENLEENN}  
In particular, the pseudogap metallic state was predicted
by the slave-boson approach prior to experimental observation;\cite{KL}
\\ \noindent
\hspace*{1ex}\textit{ii)}
The nature of fluctuations in each phase is also consistent with experiments.
For instance, the $U(1)$ gauge structure in the s-flux state accounts for the
spin excitation spectrum in underdoped HD materials.\cite{RW_SPIN}  It also
leads to the lack of well defined quasiparticles.\cite{RW_SPECTRAL}  In the
strange-metal regime $SU(2)$ gauge fluctuations are responsible for the
observed non-Fermi liquid behavior;
\\ \noindent
\hspace*{1ex}\textit{iii)}
MF states have properties that survive projection even in the presence of long
range $U(1)$ gauge fields.  Take the s-flux state, which includes both
staggered current and d-wave pairing fluctuations, as an example.  Remarkably,
in Ref. \onlinecite{IVANOVLW} staggered vorticity correlations were reported to
emerge from Gutzwiller projecting the dSC wave function.  There is also
numerical evidence that $Z_2$ MF states lead to fractionalized phases after
performing Gutzwiller projection.\cite{IVANOVSENTHIL}

\section{\label{sec:p_d} MF phase diagrams}

To implement the self-consistent MF equations for the $tt'J$ model, 
the uniform ansatz considered
in Ref. \onlinecite{WENLEENN} must be extended to include the NNN MF parameters
$U_{i,i+\hat{x}\pm\hat{y}}$.
Here we only consider translation invariant ansats\"e which do not
break any symmetry.  
In our MF calculation we use the following ansatz which properly comprises 
both the s-flux and dSC states: 
\begin{eqnarray}
U_{i,i+\hat{x}} &=& \chi \tau^{3} + \eta \tau^{1}, \nonumber\\
U_{i,i+\hat{y}} &=& \chi \tau^{3} - \eta \tau^{1},  \nonumber\\
U_{i,i+\hat{x}\pm \hat{y}} &=& \gamma \tau^{3},  \nonumber\\
a_{0}^{3} &\neq & 0.
\end{eqnarray}
Depending on the
values of $\chi$, $\eta$, and $\gamma$, the above ansatz can describe phases
with the same symmetry but different quantum orders.\cite{WEN}  These phases
may contain $SU(2)$, $U(1)$, or $Z_2$ gauge fluctuations.  Some phases display
a large Fermi surface while others have only Fermi points.  These different MF
states and their labels are summarized in Table \ref{tab:labels}.

\begin{table}
\begin{ruledtabular}
\begin{tabular}{|c|c|c|c|c|}
\hline
MF & Gauge & \multicolumn{3}{c|}{MF} \\
phase  & structure & \multicolumn{3}{c|}{parameters}\\
\hline
s-flux (sf) & $U(1)$ & $\chi\neq\eta$ & $\gamma=a_{0}^{3}=\rho=0$ & $\chi,\eta\neq0$ \\
\hline
Z2 & $Z_2$ & $\chi\neq\eta$ & $\chi,\eta,\gamma,a_{0}^{3}\neq0$ & $\rho=0$ \\
\hline
dSC &  $Z_1$ & $\chi\neq\eta$ & 
\multicolumn{2}{c|}{$\chi,\eta,\gamma,a_{0}^{3},\rho\neq0$} \\
\hline
U1 & $U(1)$ & $\eta=\rho=0$ & 
\multicolumn{2}{c|}{$\chi,\gamma,a_{0}^{3}\neq0$} \\
\hline
FL & $Z_1$ & $\eta=0$ & 
\multicolumn{2}{c|}{$\chi,\gamma,a_{0}^{3},\rho\neq0$} \\
\hline
uRVB & $SU(2)$ & $\chi\neq0$ & 
\multicolumn{2}{c|}{$\eta=\gamma=a_{0}^{3}=\rho=0$} \\
\hline
$\pi$-flux ($\pi$f) & $SU(2)$ & $\chi=\eta\neq0$ & 
\multicolumn{2}{c|}{$\gamma=a_{0}^{3}=\rho=0$} \\
\hline
\end{tabular}
\end{ruledtabular}
\caption{\label{tab:labels} Labels for MF phases. $\rho$ is the density of
condensed holons. All other parameters are defined in the main text}
\end{table}

The MF phase diagrams are determined for $t=3J$ and $t'$ between $-1.5J$ and
$1.5J$.  These values are representative of the cuprates (note that $t'$ is
close to $J$ for ED samples).  Holons may condense at nonzero temperatures
due to interactions or to the projection constraint (as proposed in
Ref. \onlinecite{WENLEENN}).  However, at MF level holons are free and
\textit{conventional} holon condensation only occurs at $T=0$.  In our
calculations we take $T>0$ and, hence, only consider the
\textit{unconventional} mechanism.

\begin{figure}
\includegraphics[width=0.48\textwidth]{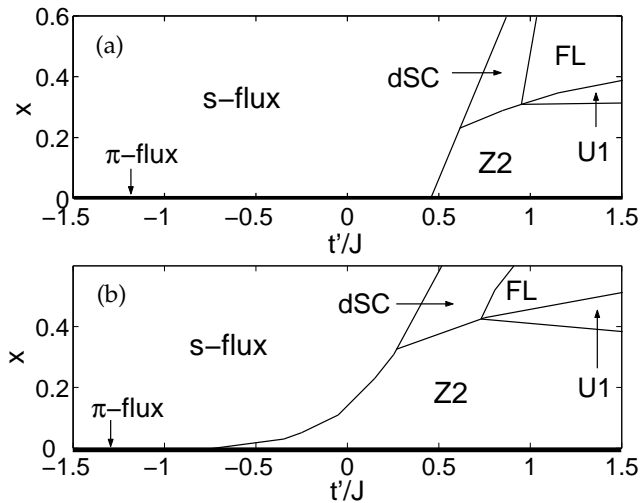}
\caption{\label{fig:xt'}(a) ($t'$,x) phase diagram for $H_{MF}$ at $T=0^{+}$. 
(b) ($t'$,x) phase diagram at $T=0^{+}$ after including $H'$ and $H''$}
\end{figure}

The resulting $(t',x)$ phase diagram for $T=0^{+}$ is shown in
Fig.\ref{fig:xt'}(a).
For $t' \lesssim 0.5J$ the s-flux state survives all the way to $T=0^{+}$ --
it becomes the dSC state for $T=0$.  However, for $t' \gtrsim 0.5J$ new MF
states are obtained.  In the underdoped regime, where the $SU(2)$ theory was
proposed to be relevant, the Z2 phase emerges.  When $T=0$ holons condense in
the band bottom and the Z2 state changes into the dSC phase.  For high enough
doping we find the dSC state even at nonzero temperature -- resulting from
projection constraint driven condensation.  As the pseudogap closes
($\eta\rightarrow 0$) a state with $U(1)$ gauge structure and gapless spinons
arises (U1).  Upon holon condensation it becomes a Fermi liquid (FL).

\begin{figure}
\includegraphics[width=0.48\textwidth]{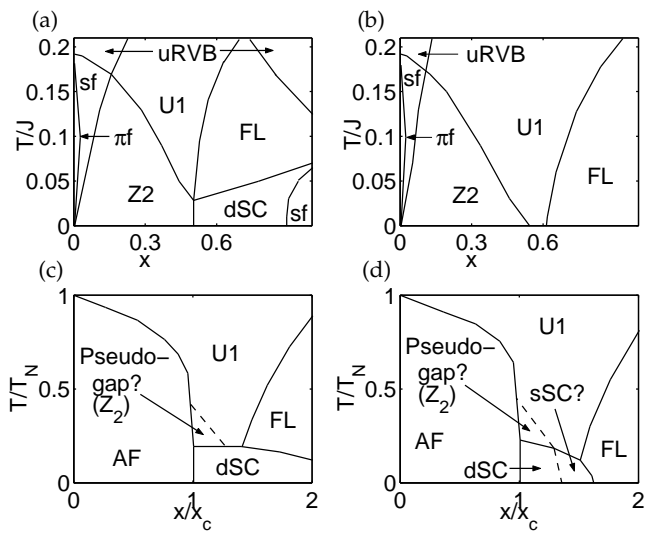}
\caption{\label{fig:xT_p08_p12}(x,T) MF phase diagrams for (a) $t'=0.8J$ and (b) $t'=1.2J$. 
In (c) and (d) \textit{qualitative} phase diagrams, based on the MF results
from (a) and (b) respectively, are proposed.  These include the experimentally
observed AF phase and finite temperature holon condensed phase due to holon 
interaction.  The dashed line describes the transition into the Z2 state
in case it is not completely absorbed by the AF phase.  Above $T_c$ the Z2
state describes pseudogap behavior. In (d) this transition may describe the
reported transition from dSC to an extended s-wave SC (sSC) phase in 
Pr$_{2-x}$Ce$_x$CuO$_{4-y}$ and La$_{2-x}$Ce$_x$CuO$_{4-y}$
(Ref. \onlinecite{PAIRING}).}

\end{figure}

For $t' \lesssim 0.5J$ we obtain the $(x,T)$ phase diagram reported 
previously.\cite{WENLEENN} The MF finite temperature phase diagrams 
are illustrated for $t'=0.8J$ and $t'=1.2J$ in Figs.  \ref{fig:xT_p08_p12}(a)
and \ref{fig:xT_p08_p12}(b) respectively.  
As, by construction, the AF phase is absent, the $Z_2$-linear state appears as
the dominant phase at low doping.  ED samples can have three different normal
metallic states: 
\textit{(a)}
a Z2 state with a $d$-wave pseudogap, true
spin-charge separation and a nontrivial topological
order;\cite{RS9173,WENsrvb,SF0050}
\textit{(b)}
a strange metal (the U1 phase) with a large Fermi surface and a $U(1)$ gauge
interaction, and
\textit{(c)}
a FL phase.

The above discussion of the MF phase diagrams is qualitatively correct even
after including the dropped $H'$ term. In terms of spinons and holons, $H'$
has the form 
\begin{eqnarray}
H' &=& -\frac{J}{2} \sum_{i} ( \Psi_{i}\sd \vec{\tau} \Psi_{i}) 
( h_{i}\sd \vec{\tau} h_{i} ) \nonumber\\
&& -\frac{J}{16} \sum_{i} \sum_{\hat{u} =
\hat{x},\hat{y}} h_{i}\sd h_{i} h_{i+\hat{u}}\sd h_{i+\hat{u}} + O(x^{3}),
\end{eqnarray}
(which is $\propto x^{2}$ at the MF level).  
While the second term correctly describes the
attraction between holons resulting from the background with AF fluctuations,
the first one should have the opposite sign to correctly account for the
attraction between spinons and holons caused by gauge fluctuations.  
Hence, $H'$ should be considered
together with an effective contribution arising from the gauge interaction.  
At MF
level the constraint is implemented only globally, \textit{i.e.} $\big<
\sum_{i} \vec{Q}_{i} \big> = 0$.
Including the following effective interaction, $H'' = \frac{3J}{16}
\sum_{i,\hat{u} = \hat{x},\hat{y}} (\vec{Q}_{i}-\vec{Q}_{i+\hat{u}} ) ^{2}$,
the fluctuations of $(\vec{Q}_{i})^{2}$ away from zero are reduced.  
The prefactor in $H''$ is such that the exchange constant is renormalized 
to $J_{eff} = 1.5J$
(note that Gutzwiller projection is known to increase the effective exchange 
constant \cite{ZHANG}).
$H''$ also makes holons 
more massive.  Including $H'$ and $H''$ at the MF level extends the 
region where the Z2 state emerges (Fig.\ref{fig:xt'}(b)). 
Hence gauge fluctuations stabilize the Z2 phase.
As expected, projection has a smaller effect in this phase. 

The above results support that the $Z_2$-linear spin liquid introduced in
Sec. \ref{sec:su2_mf} is a competitive state for higher values of $t'$ and low
doping levels.  In particular, the lower bound $t' \gtrsim 0.5J$ makes this
state relevant for ED cuprates.

\section{\label{sec:t'} Spin-charge separation: the role of $t'$}

The $SU(2)$ slave-boson approach is designed to address the low doping regime.
As it is clear from Fig.\ref{fig:xt'}, for low doping there are two dominant
phases, namely, the s-flux state and the Z2 state.  The former becomes
the latter as $t'$ is increased above a certain critical value.  This
parameter driven transition links two states with the same symmetries (they
both describe totally symmetric spin liquids).  However they have different
quantum orders.\cite{WEN}  This statement can be made more explicit by
emphasizing the quite different nature of their low-energy effective behavior.

In the s-flux state massless Dirac fermions (spinons) and charged bosons
(holons) interact with a long-range $U(1)$ gauge field.  The gapless nature of
these interacting excitations is protected by the quantum order.\cite{WEN}
Still, the gauge field has a drastic effect on the nature of the excitations
-- this state is in fact a manifestation of a two-dimensional Luttinger liquid 
\cite{RW_SPIN,RW_SPECTRAL,FT0103} which has no well defined quasiparticles. 

On the other hand, in the Z2 state the gauge interaction becomes short-ranged.
Hence, the (linearly dispersing) spinons and holons are well defined
quasiparticles.  For that reason there is true spin-charge separation.  The
remnant discrete $Z_2$ gauge structure reflects the topological order of the
state.\cite{WENsrvb} This topological order is related to the ground-state
degeneracy of the system when it is embedded in a manifold with nontrivial
topology.  Such degenerate ground-states are locally similar but have
different global (topological) properties.

The different gauge structure in the two states ($U(1)$ and $Z_2$) naturally
gives rise to qualitatively distinct spectra for the collective modes.  Indeed,
the transition is accompanied by the opening of a gap in the fluctuations
around the MF saddle-point.\cite{RIB_WEN}

Physically,
the transition from the s-flux state to the Z2 state corresponds to
the emergence of coherent diagonal (intrasublattice) charge carrier 
hopping -- \textit{i.e.} 
$\big< c_{i}\sd c_{i+\hat{u}} + c_{i+\hat{u}}\sd c_{i} \big>$, 
with $\hat{u} = \pm \hat{x} \pm \hat{y}$, becomes nonzero.
It results from a combination of two factors: 
\textit{(a)}
increasing $t'$ unfrustrates diagonal hopping and
\textit{(b)}
electron/hole intrasublattice hopping is not frustrated and does not
frustrate the background AF correlations (unlike intersublattice hopping).
These play a crucial role in the dynamics of charge carriers 
(and of spin degrees of freedom as well).

In the s-flux state intrasublattice hopping is depleted by the negative $t'$.
As charge carriers hop between different sublattices they interact strongly
with the background AF fluctuations.  Hence charge carrier motion and AF
correlations frustrate each other.  That is the motive underlying the
staggered current and staggered chiral spin correlations.\cite{S_CHIRAL} The
presence of such correlations in the context of the $tJ$ model was reported in
exact diagonalization studies.\cite{LEUNG}  Moreover, charge carrier density
and vorticity correlations are proportional to each other.\cite{IVANOVLW,LEUNG}
 The emerging picture consists of charge carriers of opposite
vorticity attracting each other in a background of staggered chiral spin
fluctuations.  This attraction eventually induces dSC coherence.

In the Z2 state charge carriers hop coherently along the diagonals
and the staggered current correlations decrease.
In particular, the energy difference between the dSC and 
staggered current states is seen to increase with $t'$.
Therefore, staggered current fluctuations are reduced in the Z2 state.
Exact diagonalization studies in doped $tt'J$ ladders show the same trend
-- \textit{i.e.} staggered currents are stabilized by decreasing 
$t'$.\cite{TSUTSUI}
According to our calculations, spinon pairing ($\eta$) is also reduced 
by the emergence of coherent diagonal hopping (therefore, charge carrier
pairing may be expected to decrease as well).
This is further consistent with the fact that the $Z_2$-vortex gap is 
physically connected to the reduction of staggered chiral spin 
fluctuations.\cite{RIB_WEN,SMALL_CLUSTERS}

Topological order in the Z2 state is associated with the intrasublattice
hopping parameter $\gamma$.  Studies of Gutzwiller-projected wave functions in
Ref. \onlinecite{IVANOVSENTHIL} further corroborate the importance of the NNN
hopping term (\textit{i.e.} a non-vanishing $\gamma$) in realizing true
fractionalization.
Consequently, for underdoped samples, the $Z_2$-vortex gap in this
fractionalized metallic pseudogap phase is set by $\gamma$ which is
$\propto x$.  This is
qualitatively different from the suggestion of
Senthil and Fisher where the $Z_2$-vortex gap is set by the pseudogap scale
($\eta$).\cite{Z2CHECK_TH}

The arguments above support and delineate how the parameter $t'$ 
induces spin-charge separation.
However, to fit experimental band structure, an extra hopping
term ($t''$) must be included.\cite{KIM_TSUTSUI}
This requires  an extension of the ansats\"e considered in our analysis.
Nevertheless, in a tight-binding model with $t>0$,
$t'>0$ and $t''<0$ (as it is the case for ED cuprates), $t''$ is seen to
frustrate hopping along the lattice bonds while unfrustrating hopping along
the diagonals  -- in this sense enhancing the effect of $t'$. 

\section{\label{sec:application}Application to ED cuprates}

As mentioned before, in ED cuprates the AF state is very 
stable.
Such stability can be addressed within the $tt'J$ model framework.
Indeed, an increasing $t'$ unfrustrates intrasublattice hopping and renders 
the charge carriers less effective in frustrating the background AF 
correlations.\cite{TOHYAMA_GOODING}
However, our MF results suggest that, if AF order can be destroyed by tuning
parameters in the Hamiltonian, the resulting spin liquid may contain a
pseudogap and correspond to the Z2 state.

A distinctive experimental feature of the Z2 phase is the 
presence of a $d$-wave pseudogap.\cite{AF_GAP}
Well, in the different Nd$_{2-x}$Ce$_x$CuO$_{4-y}$, 
Pr$_{2-x}$Ce$_x$CuO$_{4-y}$ (PCCO), and La$_{2-x}$Ce$_x$CuO$_{4-y}$ (LCCO) 
SC thin films, the SC gap 
($\sim 5-10 meV$) is reported to survive in the magnetic field driven 
normal state.\cite{KLEEFISCH_BISWAS_ALFF}
The value of the normal state gap is comparable to the SC gap and decreases
with doping.  Moreover, the low temperature normal state in PCCO SC crystals
was reported to violate the  Wiedemann-Franz law.\cite{HILL}  This provides
evidence for a non-Fermi liquid state compatible with spin-charge separation.
According to our MF results, the aforementioned normal state may realize the
Z2 state. On the other hand, if the Z2 state is to exist above $T_c$ it
probably only covers a small fraction of the phase diagram due to the
robust AF phase (Figs.\ref{fig:xT_p08_p12}(c) and 
\ref{fig:xT_p08_p12}(d)).\cite{FLIQUID}

Despite still disputed, there are a number of experiments reporting dSC order
in ED samples.\cite{TSUEI}  In our picture, the $d_{x^2-y^2}$ gap in the SC
phase is inherited from spinon pairing in the Z2 state.  Recently, however, a
SC pairing symmetry transition across optimal doping, from d-wave to extended
s-wave, was reported in both LCCO and PCCO.\cite{PAIRING}  We suggest the
extended s-wave pairing may result in the FL state 
(Fig.\ref{fig:xT_p08_p12}(d)) from exchange of collective
modes -- as the interaction between the Fermi
sea and the massive gauge-fields raises an instability in the s-wave pairing
channel.\cite{PATRICKLEE}  We remark that this mechanism is unrelated to the
one underlying dSC.

\section{\label{sec:conclusion}Conclusions}

In this paper we use a self-consistent $SU(2)$ slave-boson approach
to study the role of NNN hopping in the context of the $tt'J$ model.  We find
that $t'$ induces a transition between the s-flux state (an algebraic spin
liquid \cite{RW_SPIN,RW_SPECTRAL}) and the Z2 state (a spin liquid with 
true spin-charge separation).
This transition is interpreted as the emergence of coherent intrasublattice
hopping in the Z2 state. The parameter $t'$ is also found to decrease 
staggered current and staggered chiral spin fluctuations, as well as to 
reduce pairing between charge carriers.

The range of parameters involved leads us to propose that a pseudogap
metallic state in ED cuprates is likely to be described by a $Z_2$-linear
phase.  In contrast, the pseudogap state for HD samples is found to be
described only by the s-flux phase with $U(1)$ gauge interaction.

Realizing fractionalized states in real systems is a major challenge.  The
present work suggests that the search for a pseudogap metallic state in ED
cuprate samples may achieve such a goal.

We would like to acknowledge many discussions with W. Rantner. TCR
was partially supported by the Praxis XXI Grant No. BD/19612/99 (Portugal).
TCR and XGW were also supported by NSF Grant No. DMR--01--23156
and by NSF-MRSEC Grant No. DMR--02--13282.

\newcommand*{\PR}[1]{Phys.\ Rev.\ {\textbf {#1}}}
\newcommand*{\PRL}[1]{Phys.\ Rev.\ Lett.\ {\textbf {#1}}}
\newcommand*{\PRB}[1]{Phys.\ Rev.\ B {\textbf {#1}}}
\newcommand*{\PTP}[1]{Prog.\ Theor.\ Phys.\ {\textbf {#1}}}
\newcommand*{\MPL}[1]{Mod.\ Phys.\ Lett.\ {\textbf {#1}}}
\newcommand*{\JPC}[1]{Jour.\ Phys.\ C {\textbf {#1}}}
\newcommand*{\RMP}[1]{Rev.\ Mod.\ Phys.\ {\textbf {#1}}}
\newcommand*{\RPP}[1]{Rep.\ Prog.\ Phys.\ {\textbf {#1}}}
\newcommand*{\PHY}[1]{Physics {\textbf {#1}}}
\newcommand*{\ZP}[1]{Z.\ Phys.\ {\textbf {#1}}} 
\newcommand*{\JETP}[1]{Sov.\ Phys.\ JETP Lett.\ {\textbf {#1}}}
\newcommand*{\PLA}[1]{Phys.\ Lett.\ A {\textbf {#1}}}
\newcommand*{\AP}[1]{Adv.\ Phys.\ {\textbf {#1}}}
\newcommand*{\JLTP}[1]{J.\ Low Temp.\ Phys.\ {\textbf {#1}}}
\newcommand*{\SC}[1]{Science\ {\textbf {#1}}}
\newcommand*{\NA}[1]{Nature\ {\textbf {#1}}}
\newcommand*{\CMAT}[1]{cond-mat/{#1}}
\newcommand*{\JPSJ}[1]{J.\ Phys.\ Soc.\ Jpn.\ {\textbf {#1}}}
\newcommand*{\PC}[1]{Physica.\ C {\textbf {#1}}}
\newcommand*{\JPCS}[1]{J.\ Phys.\ Chem.\ Solids.\ {\textbf {#1}}}
\newcommand*{\APNY}[1]{Ann.\ Phys.\ (N.Y.) {\textbf {#1}}}
\newcommand*{\SSC}[1]{Solid State Commun.\ {\textbf {#1}}}
\newcommand*{\SST}[1]{Supercond.\ Sci.\ Technol.\ {\textbf {#1}}}


\begin{thebibliography}{99}
\bibitem{SMALL_CLUSTERS} Results based on numerical calculations on ladder systems show that hole binding is favored by an increasing $t'$ (Refs. 
\onlinecite{TSUTSUI} and \onlinecite{WHITE_MARTINS}).
The sensitiveness to the effective dimensionality of the system under analysis is reflected in Fig.12 of Ref. \onlinecite{G_V_LEUNG}.
In Ref. \onlinecite{WHITE_MARTINS} some qualitative arguments on small clusters were put forth to motivate the aforementioned results. 
These differ from the picture we propose. In the present approach the pairing parameter $\eta$ (which is not the superconducting order parameter obtained from the electron d-wave pair field correlation function) emerges in the spin sector due to the highly frustrated antiferromagnetic
 background with non-local correlations. 
As suggested above such pairing is enhanced by staggered chiral spin fluctuations.
Our results support that increasing $t'$ reduces these fluctuations and facilitates single charge carrier intrasublattice hopping.
\bibitem{AF_GAP} There is a large gap, evolving from
$\sim 400 meV$ for $x=0.04$ to $\sim 100 meV$ for $x=0.15$, in the stable AF
phase in ED samples of Nd$_{2-x}$Ce$_x$CuO$_{4-y}$ according to ARPES 
(Ref. \onlinecite{ARPES_ONE_TWO})
and optical conductivity (Ref. \onlinecite{ONOSE}) measurements.  
It results from a
$(\pi,\pi)$ scattering mechanism (Ref. \onlinecite{ARPES_THREE}) 
and is not the pseudogap in the Z2 state.
\bibitem{FLIQUID} As noted above, in the absence of the Z2 pseudogap regime above $T_c$ our MF results indicate that the normal state can be described by either a strange metal (U1 phase) or a Fermi liquid.
\end{thebibliography}
\end{document}